\documentclass[aps, reprint, amsmath, amssymb, amsfonts, floatfix, intlimits,superscriptaddress]{revtex4-1}
\pdfoutput=1
\usepackage{graphicx}
\usepackage{dcolumn}
\usepackage{bm}
\usepackage{color} 
\usepackage{float}
\usepackage{gensymb}
\usepackage{amsmath}
\usepackage{amssymb}

\definecolor{deepBlue}{RGB}{48, 71, 210}
\definecolor{myGreen}{RGB}{200, 38, 216}
\definecolor{APSBlue}{RGB}{46, 48, 146}
\usepackage{url}
\usepackage{color}
\usepackage[colorlinks = true,
	linkcolor = blue,
	urlcolor  = blue,
	citecolor = blue,
	anchorcolor = blue]{hyperref}

\begin{document}
	
	\title{Magic Angle Spectroscopy}
	
	\affiliation{
		Department of Physics, Columbia University, New York, New York 10027, United States\looseness=-1}
	\affiliation{
		Dahlem Center for Complex Quantum Systems and Fachbereich Physik, Freie Universit$\ddot{a}$t Berlin, 14195 Berlin, Germany\looseness=-1}
	\affiliation{
		Max Planck Institute for the Structure and Dynamics of Matter, Luruper Chaussee 149, 22761 Hamburg, Germany\looseness=-1}
	\affiliation{
		Department of Applied Physics and Applied Mathematics, Columbia University, New York, NY, USA\looseness=-1}
	\affiliation{
		National Institute for Materials Science, 1-1 Namiki, Tsukuba 305-0044, Japan\looseness=-1}
	\affiliation{
		Department of Mechanical Engineering, Columbia University, New York, NY, USA\looseness=-1}
	\affiliation{
		Center for Computational Quantum Physics (CCQ), The Flatiron Institute, 162 Fifth Avenue, New York, NY 10010, USA\looseness=-1}

	\author{Alexander Kerelsky}
	\affiliation{
		Department of Physics, Columbia University, New York, New York 10027, United States\looseness=-1}
	\author{Leo McGilly}
	\affiliation{
		Department of Physics, Columbia University, New York, New York 10027, United States\looseness=-1}
	\author{Dante M. Kennes}
	\affiliation{
		Dahlem Center for Complex Quantum Systems and Fachbereich Physik, Freie Universit$\ddot{a}$t Berlin, 14195 Berlin, Germany\looseness=-1}
	\author{Lede Xian}
	\affiliation{
		Max Planck Institute for the Structure and Dynamics of Matter, Luruper Chaussee 149, 22761 Hamburg, Germany\looseness=-1}
	\author{Matthew Yankowitz}
	\affiliation{
		Department of Physics, Columbia University, New York, New York 10027, United States\looseness=-1}
	\author{Shaowen Chen}
	\affiliation{
		Department of Physics, Columbia University, New York, New York 10027, United States\looseness=-1}
	\affiliation{
		Department of Applied Physics and Applied Mathematics, Columbia University, New York, NY, USA\looseness=-1}
	\author{K. Watanabe}
	\affiliation{
		National Institute for Materials Science, 1-1 Namiki, Tsukuba 305-0044, Japan\looseness=-1}
	\author{T. Taniguchi}
	\affiliation{
		National Institute for Materials Science, 1-1 Namiki, Tsukuba 305-0044, Japan\looseness=-1}
	\author{James Hone}
	\affiliation{
		Department of Mechanical Engineering, Columbia University, New York, NY, USA\looseness=-1}
	\author{Cory Dean}
	\affiliation{
		Department of Physics, Columbia University, New York, New York 10027, United States\looseness=-1}
	\author{Angel Rubio}
	\altaffiliation[Correspondence to: ]{
		\href{mailto:angel.rubio@mpsd.mpg.de}{angel.rubio@mpsd.mpg.de}}
	\affiliation{
		Max Planck Institute for the Structure and Dynamics of Matter, Luruper Chaussee 149, 22761 Hamburg, Germany\looseness=-1}
	\affiliation{
		Center for Computational Quantum Physics (CCQ), The Flatiron Institute, 162 Fifth Avenue, New York, NY 10010, USA\looseness=-1}
	\author{Abhay N. Pasupathy}
	\altaffiliation[Correspondence to: ]{
		\href{mailto:apn2018@columbia.edu}{apn2018@columbia.edu}}
	\affiliation{
		Department of Physics, Columbia University, New York, New York 10027, United States\looseness=-1}
	
	\date{\today}
	\begin{abstract}
		The electronic properties of heterostructures of atomically-thin van der Waals (vdW) crystals can be modified substantially by Moir\'e superlattice potentials arising from an interlayer twist between crystals. Moir\'e-tuning of the band structure has led to the recent discovery of superconductivity and correlated insulating phases in twisted bilayer graphene (TBLG) near the so-called ``magic angle'' of $\sim$1.1$^\circ$, with a phase diagram reminiscent of high T$_c$ superconductors. However, lack of detailed understanding of the electronic spectrum and the atomic-scale influence of the Moir\'e pattern has so far precluded a coherent theoretical understanding of the correlated states. Here, we directly map the atomic-scale structural and electronic properties of TBLG near the magic angle using scanning tunneling microscopy and spectroscopy (STM/STS). We observe two distinct van Hove singularities (vHs) in the LDOS which decrease in separation monotonically through 1.1$^\circ$ with the bandwidth ($t$) of each vHs minimized near the magic angle. When doped near half Moir\'e band filling, the conduction vHs shifts to the Fermi level and an additional correlation-induced gap splits the vHs with a maximum size of 7.5meV. We also find that three-fold (C$_3$) rotational symmetry of the LDOS is broken in doped TBLG with a maximum symmetry breaking observed for states near the Fermi level, suggestive of nematic electronic interactions. The main features of our doping and angle dependent spectroscopy are captured by a tight-binding model with on-site ($U$) and nearest neighbor Coulomb interactions. We find that the ratio $U/t$ is of order unity, indicating that electron correlations are significant in magic angle TBLG. Rather than a simple maximization of the DOS, superconductivity arises in TBLG at angles where the ratio $U/t$ is largest, suggesting a pairing mechanism based on electron-electron interactions.
	\end{abstract}
	\keywords{}
	\maketitle
	
	Van der Waals heterostructures comprising of two monolayers with a slight rotation yield a structural Moir\'e superlattice which often induces entirely new electronic properties \cite{lopes07,wu18,Xian18}. The superlattice has a period determined geometrically by the difference in lattice vectors and has structural distortions in each layer to minimize the overall free energy of the system. The hopping between layers further modifies the band structure of the bilayer. In recent years, twisted bilayers have been produced by growth \cite{wong15}, mechanical stacking of monolayers  \cite{kim162} and even by controllable rotation \cite{palau18}. In the case of graphene, the twisted bilayer yields two copies of the Dirac band structure which cross above and below the Dirac point \cite{lopes07}. Hybridization between the layers creates two additional vHs's at these crossing points \cite{wong15,yin15,xue11,deck11,li09,kim17,brih12}. A continuum model analysis \cite{brist11} of the band structure of TBLG predicted that at a magic angle near 1.1$^\circ$ the hybridization between the layers would push the energy of the vHs's to the Dirac point while flattening their bandwidth, thus creating an entire two-dimensional region in momentum space where the states have virtually no dispersion. The low energy physics of the electrons would then be largely determined by the Coulomb interaction and lead to the possibility of new emergent many-body ground states \cite{wu18,chen18}. Indeed, recent transport measurements have shown the presence of both superconducting \cite{cao182,yank18} and insulating states \cite{cao18} at these conditions. The phase diagram is reminiscent of unconventional superconductors, but in a two-dimensional, gate-tunable material with simple chemistry. These exciting new developments indicate that small angle twisted bilayers are new model systems in condensed matter physics where control over bandwidth and interactions can be achieved using simple experimental knobs, paving a new avenue that could hold insights into unconventional superconductivity.
	
	\begin{figure*}[t]
		\includegraphics[width=\linewidth]
		{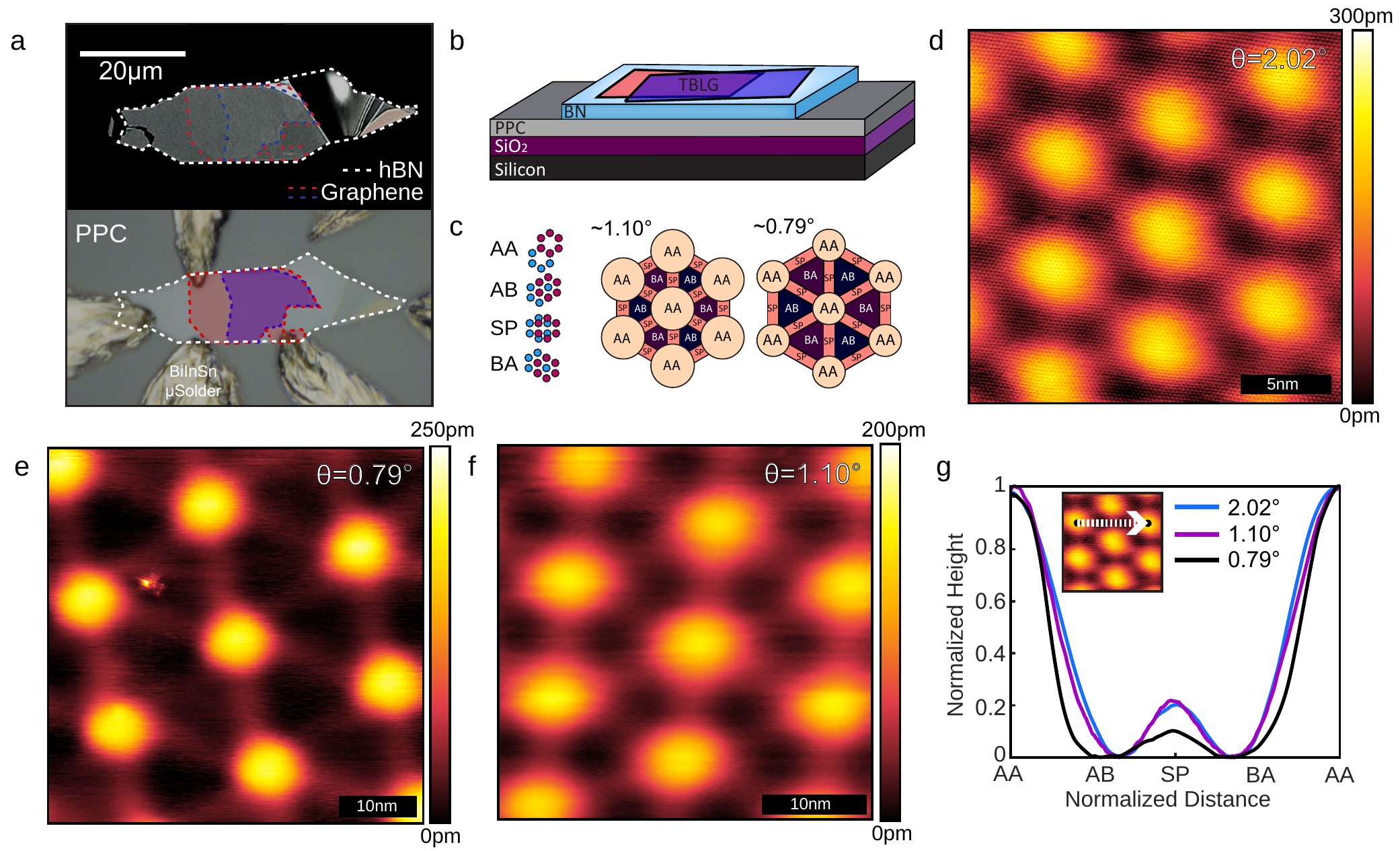}
		\caption{
			(a) Optical images of one of eight samples measured. Dashed lines highlight the layers of hBN and the two twisted monolayers of graphene. The top is mid fabrication immediately after stacking while the bottom is the final structure contacted with Field's metal. (b) Schematic of sample structure being measured. (c) Schematics of a real space Moir\'e pattern interchanging between AA, AB/BA and SP stacking. (d-f) Atomic resolution STM topographies on 2.02$^\circ$, 0.79$^\circ$ and 1.10$^\circ$ TBLG samples. Topographies were taken at 1V, 30pA, 1V, 50pA and 0.4V, 50pA respectively. (g) Normalized spatial height profiles of the AA to second nearest AA, as delineated in the cartoon, for the three angles shown in figure 1d-f.} 
	\end{figure*} 
	
	Despite rapid developments, the atomic and electronic structure of TBLG has yet to be verified. This has posed a formidable challenge for theoretical modeling of TBLG -- in particular, theory is not yet able to identify the origin of the correlated insulating phases, or whether the superconducting pairing is mediated by electronic interactions. Recent experiments have shown the importance of atomic rearrangements at small twist angles, but their effect on the electronic structure is yet to be determined \cite{kim17,yoo18,huang18}. While the presence of insulating behavior in the phase diagram is possibly a many-body effect in TBLG \cite{xu18,yuan18,thom18,po18,iso18,pahdi18,dod18,kenn18}, the role of disorder and strain is yet to be clarified. Thus, it is important to have direct measurements of the atomic structure and the low-energy electronic structure in TBLG for which STM/STS is an ideal spectroscopic tool. Here, we present direct measurement of the local angle- and doping-dependent atomic scale-structure and LDOS of near-magic angle TBLG on hBN directly measured using STM/STS at 5.7K in a homebuilt UHV-STM. To fully explore this problem, it is necessary to study TBLG samples near the magic angle on homogenous, insulating substrates with control over electrostatic doping. While previous STM works have explored TBLG, measurements were either performed on angles that are far from the magic angle \cite{wong15,kim17,huang18}, or on conducting substrates where electrostatic doping is not possible and the Coulomb interaction is screened \cite{yin15,li09,brih12}. Our samples were fabricated following the pioneering ``tear and stack'' method (see methods) that is used to fabricate transport devices \cite{kim162} where superconductivity was measured, however left uncapped for the STM measurement. An optical image of a typical sample is shown in figure 1a and a schematic in figure 1b. 
	
	\begin{figure*}[t]
		\includegraphics[width=\linewidth]
		{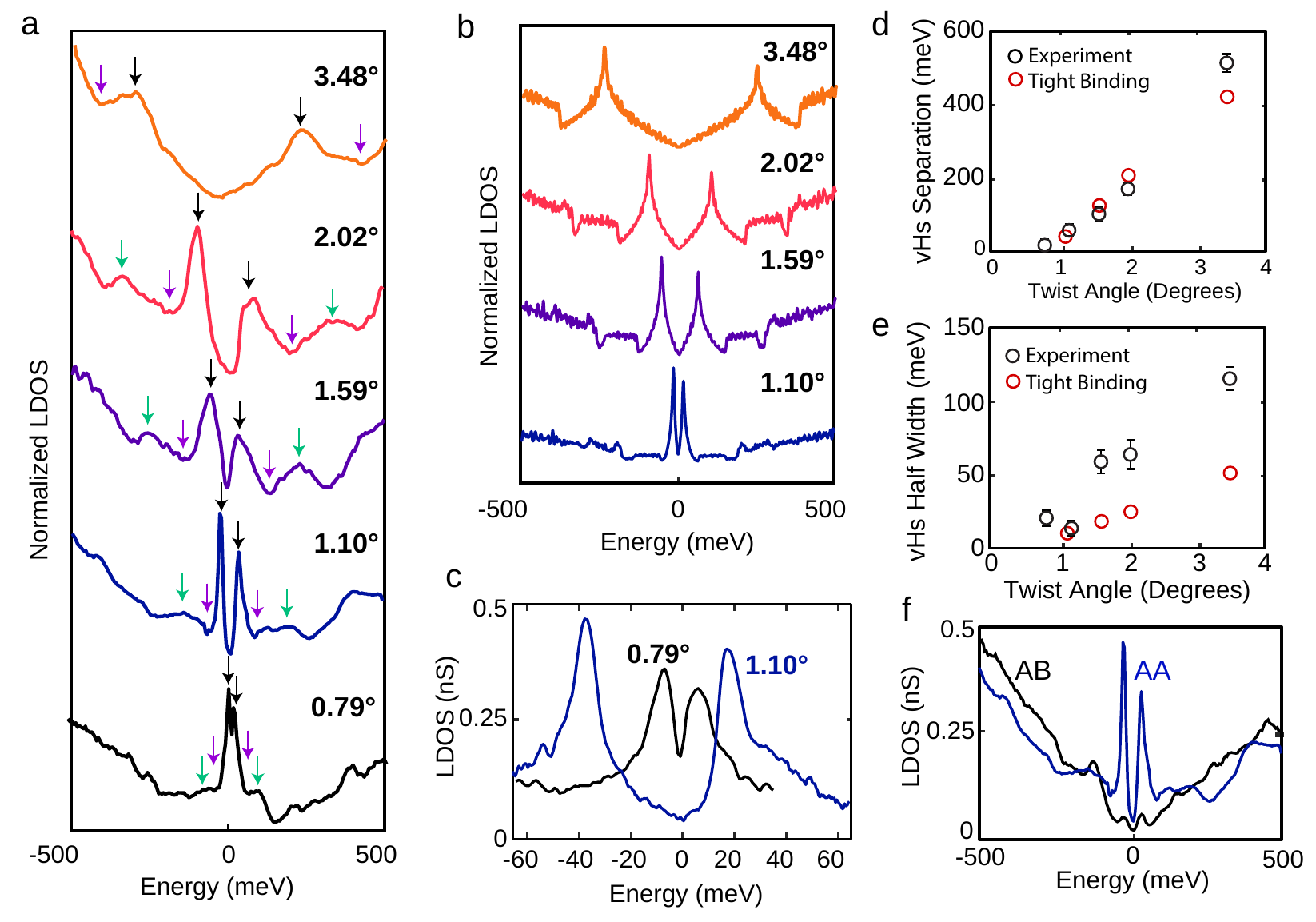}
		\caption{
			(a) STS LDOS on Moir\'e AA sites of 3.48$^\circ$, 2.02$^\circ$, 1.59$^\circ$, 1.10$^\circ$ and 0.79$^\circ$, normalized to the maximum value for each curve and vertically offset for clarity. Arrows show several features prominent consistent at all angles -- the van Hove singularities (black arrows), the first dips (purple arrows) and a second smaller peak (green arrows) previously observed. With decreasing twist angle all features shift towards the Fermi level. (b) Tight-binding calculations of the LDOS at the measured angles down to 1.10$^\circ$. (c) Higher energy resolution zoom in of STS LDOS on 1.10$^\circ$ and 0.79$^\circ$ AA sites. (d) Experimental versus tight-binding vHs separation as a function of twist angle. (e) Experimental versus tight-binding half width of each vHs as a function of twist angle. (f) STS LDOS on AA versus AB sites in 1.10$^\circ$ TBLG.} 
	\end{figure*} 
	
	Figure 1c shows the structure of a TBLG Moir\'e. Within a Moir\'e unit cell, the stacking arrangement between the two layers displays regions of AA, AB/BA (Bernal) and SP (saddle-point) stacking \cite{kim17,yoo18,huang18}. Figure 1d-f show typical atomic resolution topographic images of TBLG at various small angles as indicated. The bright regions in the STM topographies have been shown in previous STM measurements to be the AA stacking sites of the TBLG, while the dark regions are the AB/BA regions with the atomic alignment evolving accordingly. There is no signature of a TBLG-hBN Moir\'e pattern in these images. This is because we intentionally made the angle between the hBN and the TBLG large to minimize interaction between the two which can change the electronic properties \cite{hunt13}. The angle between the graphene layers can be identified by a direct measurement of the Moir\'e periods. When the two graphene lattices have no strain present, a single Moir\'e period exists in the material with period $\lambda=a/(2sin(\theta/2))$. In all of our samples as well as in previous samples studied by STM, a small amount of strain is present in one of the layers which arises at some point of the fabrication process causing the Moir\'e period along the two principal directions of the Moir\'e lattice to be slightly different. We use a more comprehensive model (supplementary S1) that accurately extracts the twist angle and the strain. The uniaxial heterostrain in our samples varies between 0.1\% and 0.7\%. Variability in strain and twist angle from expected values can heavily modify electronic transport signatures where different properties have been seen even within a single sample between different pairs of contacts \cite{yank18}.
	
	\begin{figure*}[t]
		\includegraphics[width=\linewidth]
		{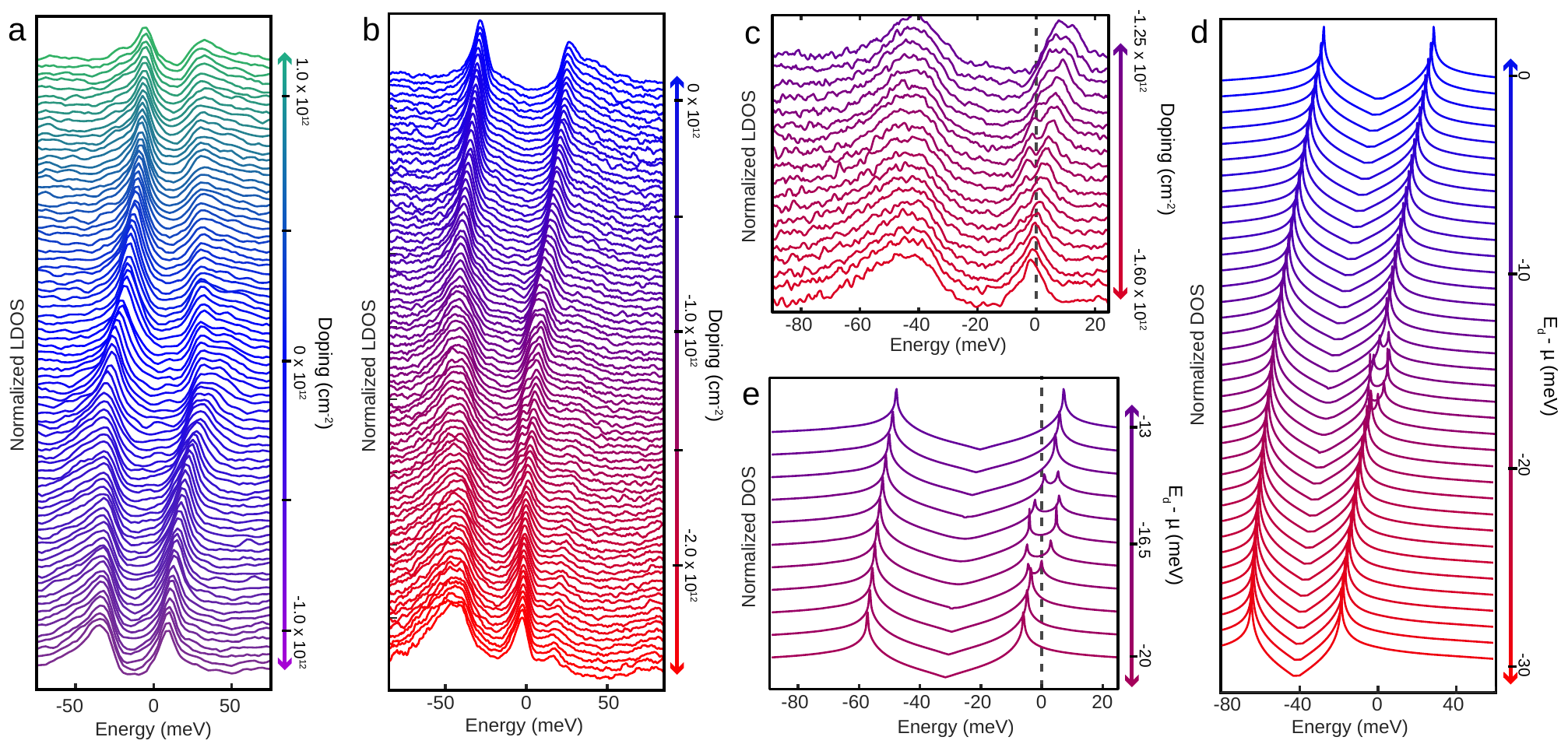}
		\caption{
			STS LDOS on a (a) 1.10$^\circ$ AA site and (b) 1.15$^\circ$ AA site as a function of doping. Curves are normalized to the maximum of the entire series and offset for clarity. Half filling of the Moir\'e cell is approximately 1.4 x 10$^{12}$ electrons/cm$^2$ for these angles. (c) Zoom in to figure 3b around half filling of the Moir\'e cell revealing a gap as the vHs crosses E$_f$. (d) Hartree-Fock mean field DOS offset as a function of chemical potential shift from neutrality, E$_d$. (e) Zoom in of figure 3(d) as around E$_d$-$\mu$=16.5meV.} 
	\end{figure*} 
	
	One important structural consideration in TBLG is the nature of the SP region that forms the interface between the AB and BA regions (see figure 1). At large twist angles ($>$4$^\circ$), the structure evolves smoothly from a AB to BA region as seen in previous experiments \cite{wong15}. At very small angles ($<$0.5$^\circ$) on the other hand, it has been observed that the material prefers to maximize the regions of AB and BA stacking, with the SP regions sharpening, producing domain walls \cite{huang18}. Angles near the magic angle thus are an interesting intermediate regime between these two extremes. Indeed, it is seen by eye that domain-wall like lines are to some degree visible in all three small angles presented in figure 1d-f. To show the evolution of the SP atomic structure as a function of angle more clearly, figure 1g shows normalized height profiles along the next nearest neighbor AA direction (dotted line in the schematic in figure 1g) for each of the three angles in figures 1d-f. These height profiles allow us to compare the apparent topographic height of the AA region versus the AB region, as well as the height and width of the transition between the AB and BA region for each angle (see supplement S2 for bias dependence of apparent topographic height). This figure shows that atomic rearrangements in the SP region are important at all of these small angles including 1.10$^\circ$, though they become especially prominent below 1$^\circ$, where the line profile shows an extended flat region of AB and BA stacking.
	
	Figure 2a shows STS measurements of the LDOS on the AA stacked regions for a series of twist angles, starting from 3.49$^\circ$ to 0.79$^\circ$. Each of these measurements have been obtained at zero external doping of the TBLG and shows a filled and an unfilled vHs that we term the valence and conduction vHs. The black arrows denote the vHs's as they shift in energy towards the Dirac point with decreasing twist angle (with other features similarly shifting) as predicted and previously shown \cite{wong15}. At the angle where superconductivity has previously been observed (1.10$^\circ$), we still clearly see two distinct peaks in the LDOS with a separation of about 55meV. At the smallest angle of 0.79$^\circ$ studied here, we see that the vHs's have nearly merged into one peak with a separation of 13meV. We can compare our experimental spectra to tight-binding calculations (see methods for details). The results of these calculations are plotted in figure 2b for angles 1.10$^\circ$ and larger and show a good match to experiment for the vHs energies as seen in figure 2d. We note that our tight-binding calculations differ from previous ones \cite{tram10} that are fitted to monolayer band structures calculated by DFT within the local density approximation (LDA) or generalized gradient approximation (GGA). It is well known that these functionals tend to strongly underestimate the Fermi velocity by about 20\% compared with experimental values, which has been shown to be a many-body correlation effect that can be corrected by means of many-body self-energy GW calculations \cite{grun08}. In our calculations, we use an intralayer hopping that is fitted to the experimental Fermi velocity for monolayer graphene \cite{kuz63} (see supplement S3 for details). At 1.10$^\circ$, the separation of the two vHs's calculated by our tight-binding model is about 41meV, comparable with the 55meV value which we measured with STS but significantly larger than those reported in other tight-binding models with DFT parameters for TBLG near 1.10$^\circ$, which are typically less than 5meV \cite{cao182,tram12}. The larger intralayer hopping in our tight-binding model also implies that the angle where the Fermi velocity vanishes is smaller than that reported in literature \cite{tram12}. We have considered several effects that can contribute to the vHs splitting observed in experiment including tip induced band bending, tip gating and the presence of heterostrain and do not believe that they significantly contribute to the measured splitting (see supplement S4).
	
	To take a closer look at the difference between experimental LDOS at 0.79$^\circ$ and 1.10$^\circ$, figure 2c shows spectroscopic measurements of the LDOS peaks over a small range in energy. These spectra clearly show that while the peaks at 0.79$^\circ$ are closer together than at 1.10$^\circ$, their energy width is larger. Plotted in figure 2e are the extracted half widths of the peaks as a function of angle. It is seen that the width of the peaks is minimized near 1.10$^\circ$, the angle around where superconductivity is observed. This indicates that it is the bandwidth of an individual vHs rather than the spacing between the vHs's that is crucial to the physics of insulating and superconducting behavior in TBLG \cite{iso18,kenn18}.
	
	\begin{figure*}[t]
		\includegraphics[width=\linewidth]
		{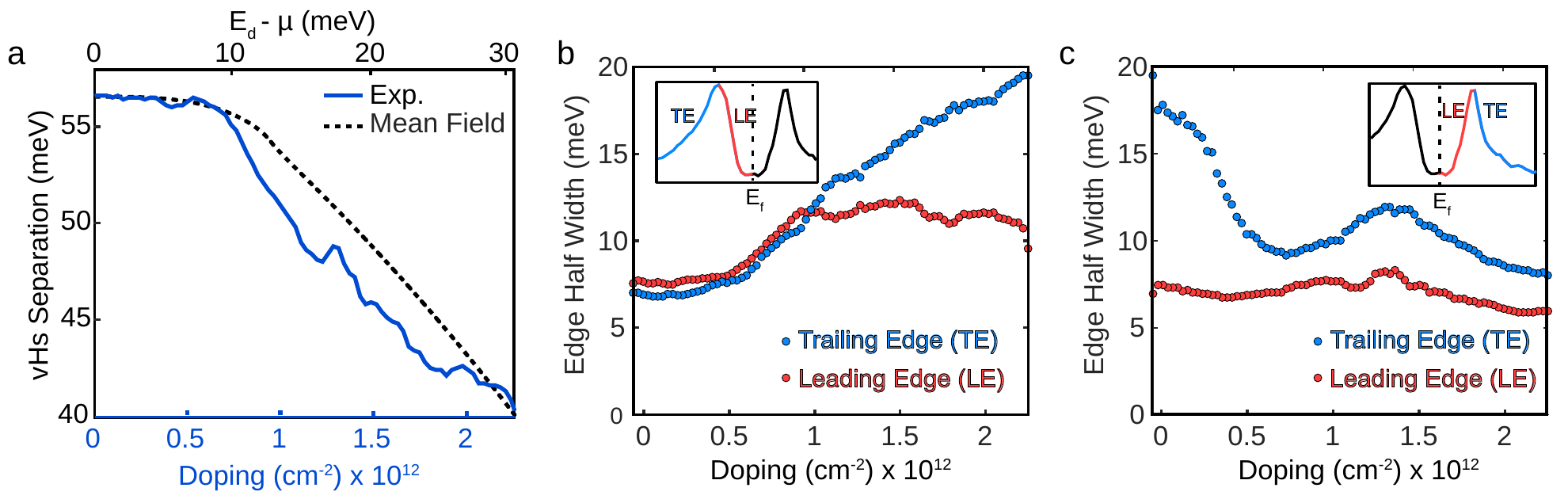}
		\caption{
			(a) Experimental vHs separation versus doping (bottom axis) and theoretical mean field vHs separation as a function of chemical potential relative to charge neutrality (top axis). (b) Half widths of the trailing and leading edge of the valence vHs as a function of doping. (c) Half widths of the trailing and leading edge of the conduction vHs as a function of doping.} 
	\end{figure*} 
	
	Having described the spectroscopic properties of 1.10$^\circ$ TBLG at zero doping we now turn to the doping dependence of the spectra on the AA site. Shown in figures 3a and 3b are sequences of spectra taken as a function of back gate voltage on two TLBG samples at 1.10$^\circ$ and 1.15$^\circ$ respectively, limited in gate voltage to where no gate leakage is observed. Due to the PPC in the structure, we estimate carrier concentration by fabricating a parallel plate capacitor and measuring the capacitance per unit area at 5.7K. From the plots, we see that the positions, shapes and separation of the vHs's are a sensitive function of the doping level. A finer set of doping dependent spectra around the region of gate voltage where the conduction vHs crosses the Fermi level is shown in figure 3c. A small gap emerges when the peak of the vHs approaches the Fermi level with a maximum peak to peak value of this gap being 7.5meV. The emergence of this gap only when the vHs is at the Fermi level is direct evidence of its many-body character. Given that the largest gap observed in transport is near half filling of the Moir\'e conduction band, it is natural to associate the gap seen here with the transport gap. In our measurement, the gap occurs when one vHs peak is at the Fermi level within experimental precision (about 1meV) which is within error of half-filling of the conduction Moir\'e band in doping (1.4 x $10^{12}$ cm$^{-2}$ for 1.15$^\circ$). In transport, additional gaps are seen at quarter filling and three-quarter filling of the Moir\'e bands \cite{yank18,cao18}. We have not seen these in spectroscopy, possibly because they are too weak at the temperature of our measurement. The magnitude of the gap observed here is however significantly bigger than the activation energy of the resistance in transport measurements. This is likely due to disorder averaging, which always produces smaller activation gaps in transport than those measured in spectroscopy \cite{xia10} motivating future lower temperature measurements.
	
	Next, we discuss the separation between the two vHs's, which is maximum at charge neutrality and reduced with doping in either direction. This behavior is reminiscent of correlation effects on the quasiparticle gap in 2D semiconductors with doping. We model this with a simple one-band model on a nearest neighbor hopping honeycomb lattice with nearest-neighbor hopping $t_0$ = 16.3meV. We include correlations via on-site and nearest neighbor repulsive interactions $U$ and $V_1$, respectively, and study the spectrum of the system in the Hartree-Fock approximation (see supplement S5). The results of the calculations are shown in figures 3d-e.  The on-site interaction $U$ open the correlation-gap when the Fermi level is tuned to a vHs. Within mean-field theory, the ordered state at the vHs filling is an antiferromagnetic state set by the nesting. We find that a value of $U$ = 4.03 meV nicely reproduces the gap seen in STS (figure 3c). The nearest neighbor interaction $V_1$ renormalizes the hopping via its Fock contribution, leading to a doping-dependent vHs splitting. We find that a value of $V_1$ = 6.26meV best reproduces the experimental dependence of splitting with doping. In figure 4a we plot the theoretical vHs separation as a function of chemical potential on top of the experimental vHs separation as a function of doping. Theory accurately captures the experimental fact that the splitting is relatively doping independent near charge neutrality but decreases strongly at high doping. The simple model shows the moderately correlated nature of magic angle TBLG with a ratio $U/t$ of order unity.
	
	\begin{figure*}[t]
		\includegraphics[width=\linewidth]
		{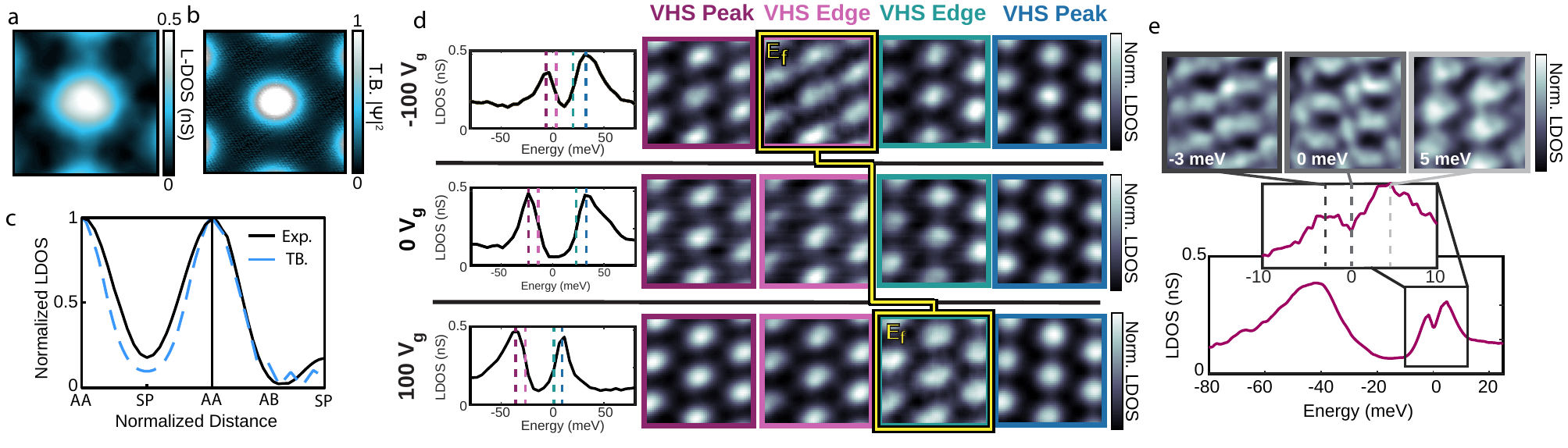}
		\caption{
			(a) STS LDOS spatial map at 50meV above E$_f$ of a Moir\'e cell (centered at AA) in the undoped TBLG at 1.10$^\circ$. (b) Probability density distribution of a single K-point wavefunction calculated using tight-binding (c) Spatial line-cuts comparing the experimental and tight-binding LDOS in the nearest and second nearest AA directions. (d) STS images at different doping levels between the two vHs peaks. The top row is at $\sim$1.2 x 10$^{12}$ holes/cm$^2$, middle is at neutrality, and bottom is at $\sim$1.2 x 10$^{12}$ electrons/cm$^2$. The averaged AA site spectra for each gating are shown in the first column, followed by a slice at the valence vHs peak, valence vHs leading edge, conduction vHs leading edge and conduction vHs peak. The yellow box and line highlights the position of the Fermi level. The breaking of C$_3$ symmetry is most apparent at the Fermi level in each case. (e) STS images when the Fermi level is at the conduction vHs peak. Images are presented at each of the split peaks flanking the correlated gap, and at the midpoint of the gap as seen from the spectrum below. A strong C$_3$ symmetry breaking is observed across the vHs at this doping, in comparison with (d) where the LDOS at the vHs peak is nearly C$_3$ symmetric.} 
	\end{figure*} 
	
	The shape of the LDOS peaks in figures 3a-b also display interesting doping dependence reminiscent of other strongly-correlated materials. Plotted in figure 4b-c are the valence and conduction vHs peak half widths on their trailing and leading edges (further and closer to the Fermi level respectively). We see that when the TBLG is undoped, the valence vHs is sharpest and most symmetric. As the conduction vHs begins to approach the Fermi level with doping, the valence vHs begins to broaden, and acquires a distinct asymmetric shape with the leading edge being much sharper than the trailing edge. On the other hand, the conduction vHs sharpens as it starts to approach the Fermi level and goes through it. The most plausible mechanism for these effects is intrinsic lifetime broadening of the states with doping. Indeed, in other strongly correlated materials such as the cuprates, such asymmetric line-shapes in photoemission spectroscopy are commonly observed and are usually attributed to correlation effects \cite{kor57}. In the case of TBLG, the lifetime broadening of the states in each of the vHs LDOS peaks is related to the number of low-energy electron-hole channels that are available for decay. Unlike a simple metal where there is a large fixed number of such decay channels present, here the density of states at the Fermi level is low until it starts entering one of the vHs's. We thus expect the lifetime of the valence band to be relatively long until the conduction band hits the Fermi level, at which point we expect a strong decrease in the lifetime. As in the cuprates, we expect this mechanism to produce incoherent excitations at energies higher than the quasiparticle itself, leading to the asymmetric line-shapes seen in experiment.
	
	We next focus on the spatial dependence of the LDOS in TBLG. Figure 5a shows an STS LDOS map at 50meV near neutrality in doping of one Moir\'e cell at 1.10$^\circ$. When undoped, LDOS maps at energies around and between the vHs's show the same contrast as the topographic image itself, with higher density of states observed on the AA sites relative to the AB sites. For comparison, we have also plotted the probability density of a single wavefunction at this energy from tight-binding in figure 5b (see methods for details). For a more accurate comparison between experiments and tight-binding calculations, we plot line profiles of the experimental and tight-binding LDOS at the same energy in figure 5c (see methods). We find that the tight-binding results systematically underestimate the LDOS intensity at the SP region and is instead more tightly confined around the AA site than experiment. We find similar behavior at other energies within the vHs peak.
	
	We now consider the evolution of the STS LDOS spatial maps as a function of doping. Shown in figure 5d are four LDOS maps at three values of doping (around -1.2 x $10^{12}$ carriers/cm$^2$, neutrality and +1.2 x $10^{12}$ carriers/cm$^2$). The four maps are at various energies between the two vHs's as indicated from the color coding in the average spectrum from each doping. The top row corresponds to a doping such that the pink boxed image is at the Fermi level, while for the bottom row the green boxed image is at the Fermi level. A close examination of the images shows that all of the LDOS maps break C$_3$ rotational symmetry to some extent, as is to be expected when some strain is present. However, one can clearly see that the maximum breaking of C$_3$ symmetry occurs at the Fermi energy. This is consistent with a scenario where the normal state has a Fermi-surface driven electronic nematic susceptibility \cite{andrad18}. We can see this even more clearly when the Fermi level is brought to the peak of the conduction vHs where the correlated gap is observed. Shown in figure 5d are three LDOS maps taken at 5meV, 0meV and -3meV at this doping, corresponding to the two newly split peaks in the spectrum and the middle of the correlation gap. In this case, an even more pronounced breaking of C$_3$ symmetry is seen in the LDOS maps. This again points to a Fermi-surface driven breaking of C$_3$ symmetry in TBLG near the magic angle and possible emergent nematic order. The question of whether a translational symmetry breaking also occurs simultaneously (for example by Fermi surface nesting \cite{kim16}, supplement S6) is still open, since samples are not uniform over large enough areas (see supplement S7) to obtain accurate Fourier space information over tens of Moir\'e unit cells. 
	
	Our spectroscopic measurements of magic angle TBLG give us new insight into the nature of the superconducting and insulating states seen in transport. The vHs separation is larger than previously thought at 1.10$^\circ$, implying that the physics of superconducting and insulating states is to be understood in the context of doping through a single vHs. Regarding superconducting order, electronic pairing mechanisms such as spin fluctuations are expected to be important when the ratio of the on-site Coulomb interaction $U$ to the bandwidth $t$ is large. On the other hand, in phonon-mediated pairing scenarios, superconducting T$_c$ is improved by lowering the Coulomb pseudopotential (proportional to U) and increasing the density of states at the Fermi level. Our spectroscopic results show that $t$ is minimized near 1.1$^\circ$; $U$ on the other hand is expected to monotonically decrease with smaller angle due to the increased real space unit cell size. Superconductivity in TBLG is thus observed when $J=U^2/t$ is maximized, hence conditions that maximize electronic rather than phonon-based pairing. In our measurements, the insulating gap appears to arise at the peak of the vHs, which is naturally associated with density wave orders rather than a real-space localization of electrons. The observed breaking of C$_3$ symmetry under these conditions is indicative of possible emergent nematic order. The interaction of such ordered states with superconductivity in TBLG remains an open problem.

	\section*{Acknowledgments}
	We thank Andrew Millis, Joerg Schmalian, Liang Fu and Rafael Fernandes for helpful discussions. This work is supported by the Programmable Quantum Materials (Pro-QM) program at Columbia University, an Energy Frontier Research Center established by the Department of Energy (grant DE-SC0019443). Equipment support is provided by the Office of Naval Research (grant N00014-17-1-2967) and Air Force Office of Scientific Research (grant FA9550-16-1-0601). Support for sample fabrication at Columbia University is provided by the NSF MRSEC program through Columbia in the Center for Precision Assembly of Superstratic and Superatomic Solids (DMR-1420634). Theoretical work was supported by the European Research Council (ERC-2015-AdG694097). The Flatiron Institute is a division of the Simons Foundation. LX acknowledges the European Union's Horizon 2020 research and innovation program under the Marie Sklodowska-Curie grant agreement No. 709382 (MODHET). DMK acknowledges funding from the Deutsche Forschungsgemeinschaft through the Emmy Noether program (KA 3360/2-1). CRD acknowledges support by the Army Research Office under W911NF-17-1-0323 and The David and Lucile Packard foundation.\\
	
	\section*{Methods: Experimental Setup}
	Our fabrication of TBLG samples follows the established ``tear and stack'' method, using PPC as a polymer to sequentially pick up hBN, half of a piece of graphene followed by the second half with a twist angle. This structure is flipped over and placed on an Si/SiO$_{2}$ chip. Directly contact is made to the TBLG via $\mu$soldering with Field's metal \cite{girit08}, keeping temperatures below 80 C during the entire process to minimize the chance of layers rotating back to Bernal stacking which happens on annealing the structures.
	
	Spectroscopy measurements are carried out using a lock-in amplifier to measure the differential conductance, with a lock-in excitation of 0.5mV p-p.
	
	\section*{Methods: Tight-binding}
	We model the twisted bilayer graphene system with the following tight-binding Hamiltonian \cite{tram10}: $\boldsymbol{\mathrm{H}}\boldsymbol{\mathrm{=}}\sum_{\boldsymbol{i},\boldsymbol{j}}{{\boldsymbol{t}}_{\boldsymbol{ij}}\left.\boldsymbol{|}\boldsymbol{i}\right\rangle \left\langle \boldsymbol{j}\boldsymbol{|}\right.},$ where $t_{ij}$ is the hopping parameter between pz orbitals at the two lattice sites, ${\boldsymbol{r}}_{\boldsymbol{i}}$ and ${\boldsymbol{r}}_{\boldsymbol{j}}$, and has the following form:	$t _ { i j } = n ^ { 2 } \gamma _ { 0 } \exp \left[ \lambda _ { 1 } \left( 1 - \frac { \left| \boldsymbol { r } _ { i } - \boldsymbol { r } _ { j } \right| } { a } \right) \right] + \left( 1 - n ^ { 2 } \right) \gamma _ { 1 } \exp \left[ \lambda _ { 2 } \left( 1 - \frac { \left| \boldsymbol { r } _ { i } - \boldsymbol { r } _ { j } \right| } { c } \right) \right]$,	where a=1.412 {\AA} is the in plane C-C bond length, c=3.36 {\AA} is the interlayer separation, n is the direction cosine of ${\boldsymbol{r}}_{\boldsymbol{i}}\boldsymbol{-}{\boldsymbol{r}}_{\boldsymbol{j}}$ along the out of plane axis (z-axis), ${\gamma }_0$ (${\gamma }_1)$ is the intralayer (interlayer) hopping parameter, and ${\lambda }_1$ (${\lambda }_2$) is the intralayer (interlayer) decay constant. This tight-binding model has been shown to reproduce the low-energy structure of TBLG calculated by density functional theory (DFT) calculations with the following value for the parameters \cite{tram10}: ${\gamma }_0=-2.7eV$, ${\gamma }_1=0.48eV$, ${\lambda }_1=3.15$ and ${\lambda }_2=7.50$. However, the Fermi velocity for monolayer graphene is usually 20\% larger than what is calculated in DFT due to correlation effects that are captured by a GW calculations \cite{grun08}. To incorporate those effects, we consider a larger intralayer hopping ${\gamma }^{'_0}=1.2 {\gamma }_0$ (the experimental parameter) as previously done \cite{wong15,brih12} (see supplement S3).
	
	For the calculation of local density of states, we employ the Lanczos recursive method \cite{wang12} to calculate the LDOS in two twisted graphene sheet in real space with a system size larger than 200nm x 200nm with an effective smearing of 1meV.
	
	For the Hartree-Fock mean-field interactions model, see supplement S5.

\bibliographystyle{apsrev4-1}
	\bibliography{tblgbibtex}

\end{document}


\setcounter{equation}{0}
		\setcounter{figure}{0}
		\setcounter{table}{0}
		\setcounter{page}{1}
		\setcounter{section}{0}
		\renewcommand{\theequation}{S\arabic{equation}}
		\renewcommand{\thefigure}{S\arabic{figure}}
	\title{Supplemental Information: Magic Angle Spectroscopy}
	
	\affiliation{
		Department of Physics, Columbia University, New York, New York 10027, United States\looseness=-1}
	\affiliation{
		Dahlem Center for Complex Quantum Systems and Fachbereich Physik, Freie Universit$\ddot{a}$t Berlin, 14195 Berlin, Germany\looseness=-1}
	\affiliation{
		Max Planck Institute for the Structure and Dynamics of Matter, Luruper Chaussee 149, 22761 Hamburg, Germany\looseness=-1}
	\affiliation{
		Department of Applied Physics and Applied Mathematics, Columbia University, New York, NY, USA\looseness=-1}
	\affiliation{
		National Institute for Materials Science, 1-1 Namiki, Tsukuba 305-0044, Japan\looseness=-1}
	\affiliation{
		Department of Mechanical Engineering, Columbia University, New York, NY, USA\looseness=-1}
	\affiliation{
		Center for Computational Quantum Physics (CCQ), The Flatiron Institute, 162 Fifth Avenue, New York, NY 10010, USA\looseness=-1}

	\author{Alexander Kerelsky}
	\affiliation{
		Department of Physics, Columbia University, New York, New York 10027, United States\looseness=-1}
	\author{Leo McGilly}
	\affiliation{
		Department of Physics, Columbia University, New York, New York 10027, United States\looseness=-1}
	\author{Dante M. Kennes}
	\affiliation{
		Dahlem Center for Complex Quantum Systems and Fachbereich Physik, Freie Universit$\ddot{a}$t Berlin, 14195 Berlin, Germany\looseness=-1}
	\author{Lede Xian}
	\affiliation{
		Max Planck Institute for the Structure and Dynamics of Matter, Luruper Chaussee 149, 22761 Hamburg, Germany\looseness=-1}
	\author{Matthew Yankowitz}
	\affiliation{
		Department of Physics, Columbia University, New York, New York 10027, United States\looseness=-1}
	\author{Shaowen Chen}
	\affiliation{
		Department of Physics, Columbia University, New York, New York 10027, United States\looseness=-1}
	\affiliation{
		Department of Applied Physics and Applied Mathematics, Columbia University, New York, NY, USA\looseness=-1}
	\author{K. Watanabe}
	\affiliation{
		National Institute for Materials Science, 1-1 Namiki, Tsukuba 305-0044, Japan\looseness=-1}
	\author{T. Taniguchi}
	\affiliation{
		National Institute for Materials Science, 1-1 Namiki, Tsukuba 305-0044, Japan\looseness=-1}
	\author{James Hone}
	\affiliation{
		Department of Mechanical Engineering, Columbia University, New York, NY, USA\looseness=-1}
	\author{Cory Dean}
	\affiliation{
		Department of Physics, Columbia University, New York, New York 10027, United States\looseness=-1}
	\author{Angel Rubio}
	\altaffiliation[Correspondence to: ]{
		\href{mailto:angel.rubio@mpsd.mpg.de}{angel.rubio@mpsd.mpg.de}}
	\affiliation{
		Max Planck Institute for the Structure and Dynamics of Matter, Luruper Chaussee 149, 22761 Hamburg, Germany\looseness=-1}
	\affiliation{
		Center for Computational Quantum Physics (CCQ), The Flatiron Institute, 162 Fifth Avenue, New York, NY 10010, USA\looseness=-1}
	\author{Abhay N. Pasupathy}
	\altaffiliation[Correspondence to: ]{
		\href{mailto:apn2018@columbia.edu}{apn2018@columbia.edu}}
	\affiliation{
		Department of Physics, Columbia University, New York, New York 10027, United States\looseness=-1}

	\date{\today}
	
	\maketitle

	\onecolumngrid
	\section*{S1: Moir\'e Wavelength under Uniaxial Heterostrain}
	In general, one or both of the graphene lattices that make up the twisted bilayer can be under strain that is produced during the fabrication process. Strain that is common to both lattices is termed homostrain, while a differential strain between the two lattices is called heterostrain. The presence of homostrain of a certain percentage results in a change of the Moir\'e wavelength by the same percentage along the homostrain direction. Given the typical sub-percent strains observed in experiment, the effect of homostrain on Moir\'e wavelengths can therefore be neglected (as they cannot produce significant differences in Moir\'e wavelength). Heterostrain on the other hand has a significant effect on the Moir\'e wavelengths. In what follows, we consider the effect of uniaxial heterostrain on the Moir\'e lattice, which we find fits all of the experimental data obtained so far. We first formulate the Moir\'e pattern for the unstrained case. Let k$_1$, k$_2$, and k$_3$ be the reciprocal wavevectors of one lattice with k$_1$ aligned along the k$_x$ axis such that: 
	
	\begin{equation}
	\boldsymbol {k} _ { 1 } = \left( \begin{array} { l } { k } \\ { 0 } \end{array} \right) , \boldsymbol {k} _ { 2 } = \left( \begin{array} { c } { \cos ( 60 ) k } \\ { \sin ( 60 ) k } \end{array} \right) , \boldsymbol {k} _ { 3 } = \left( \begin{array} { c } { \cos ( 120 ) k } \\ { \sin ( 120 ) k } \end{array} \right) \text
	 { and } k = \frac { 4 \pi } { \sqrt { 3 } a _ { 0 } }
	\end{equation}
	
	This lattice is shown on the left side of figure S1a. Next we create a second lattice at a small twist angle $\theta$ as shown in the center of figure S1a.
	
	\begin{equation}
	\boldsymbol { k } _ { 1 } ^ { \prime } = \boldsymbol { R } \left( \boldsymbol { \theta } _ { T } \right) \boldsymbol { k } _ { 1 } , \boldsymbol { k } _ { 2 } ^ { \prime } = \boldsymbol { R } \left( \boldsymbol { \theta } _ { T } \right) \boldsymbol { k } _ { 2 } , \boldsymbol { k } _ { 3 } ^ { \prime } = \boldsymbol { R } \left( \boldsymbol { \theta } _ { T } \right) \boldsymbol { k } _ { 3 } \text { and } \boldsymbol { R } \left( \boldsymbol { \theta } _ { T } \right) = \left( \begin{array} { c c } { \cos \left( \theta _ { T } \right) } & { - \sin \left( \theta _ { T } \right) } \\ { \sin \left( \theta _ { T } \right) } & { \cos \left( \theta _ { T } \right) } \end{array} \right)
	\end{equation}
	
	The Moir\'e wavevectors are the differences between the rotated wavevectors and the nonrotated wavevectors as can be seen on the right side of figure S1a. Thus:
	
	\begin{equation}
	\boldsymbol { K } _ { 1 } = \boldsymbol { k } _ { 1 } ^ { \prime } - \boldsymbol { k } _ { 1 } = \left( \begin{array} { c } { k - \mathrm { kcos } \left( \theta _ { T } \right) } \\ { - k \sin \left( \theta _ { T } \right) } \end{array} \right)
	\end{equation}
	
	With some algebra one can find that:
	
	\begin{equation}
	\boldsymbol{\left | M _ { 1 } \right|} = \frac { a _ { 0 } } { 2 \sin \left( \frac { \theta _ { T } } { 2 } \right) }
	\end{equation}
	
	Next, we consider uniaxial heterostrain on one layer in reciprocal space. To apply strain to one layer in the k$_x$ direction, we simply apply a strain matrix to one of the two lattice vectors (we choose the unrotated lattice for simplicity)
	
	\begin{figure*}[t]
		\includegraphics[width=\linewidth]
		{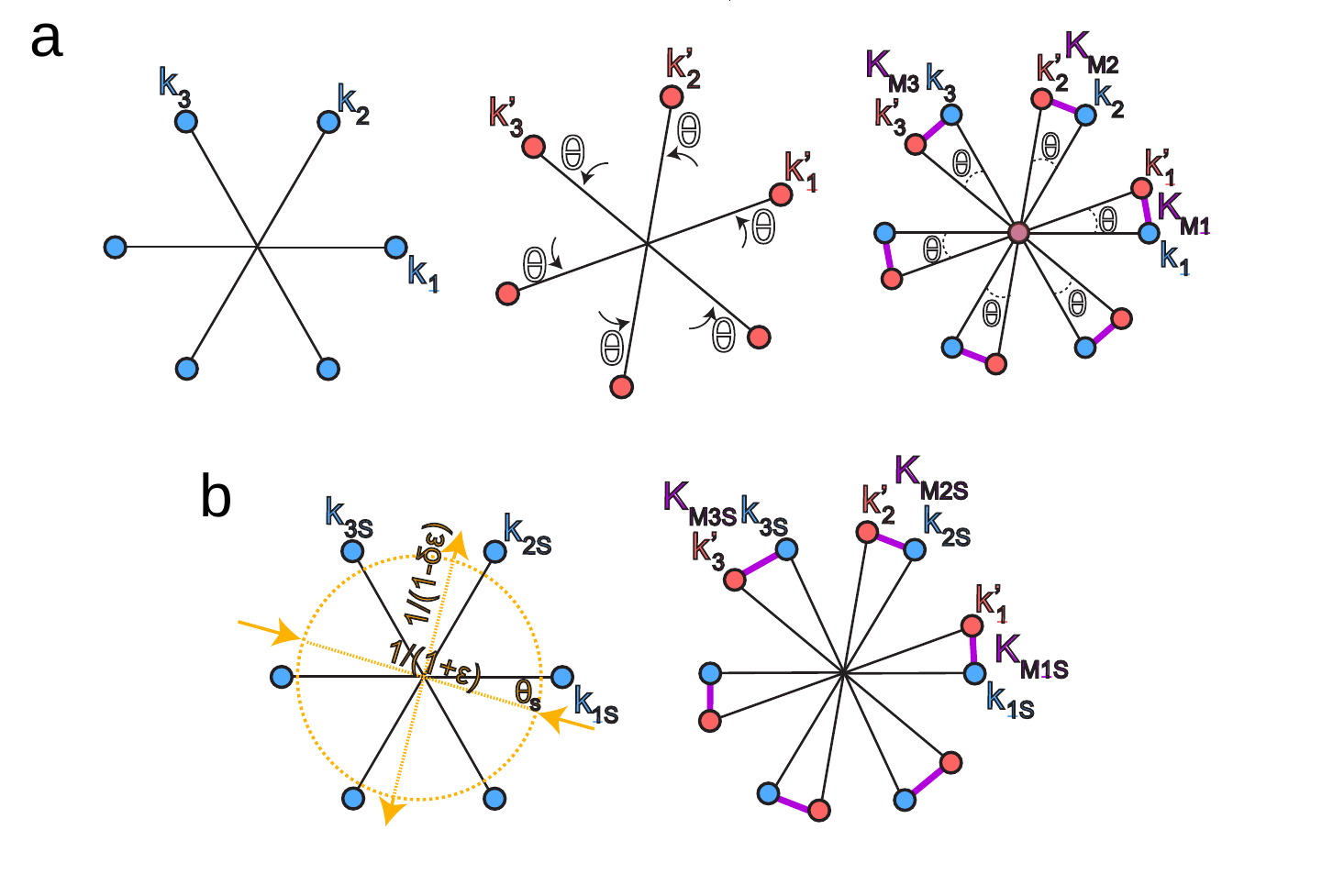}
		\caption{
			(a) Typical unstrained reciprocal space Moir\'e picture -- one layer is rotated with respect to the other producing a new periodicity characterized by the wavevectors connecting the individual layers’  reciprocal lattice vectors. (b) Uniaxial strain is applied to one layer slightly modifying the Moir\'e wavelengths. } 
	\end{figure*} 
	
	\begin{equation}
	\mathbf { E } ( \boldsymbol { \epsilon } ) = \left( \begin{array} { c c } { \frac { \mathbf { 1 } } { 1 + \epsilon } } & { \mathbf { 0 } } \\ { \mathbf { 0 } } & { \frac { \mathbf { 1 } } { \mathbf { 1 } - \delta \epsilon } } \end{array} \right)
	\end{equation}
	
	Where $\epsilon$ is the strain percentage and $\delta$ is the Poisson ratio of the material (estimated around 0.16 for graphene). If instead of the k$_x$ direction, the strain is applied at an arbitrary angle $\theta_s$ to the x axis, this can be achieved by the matrix
	
	\begin{equation}
	\mathbf { S } \left( \boldsymbol { \theta } _ { \mathbf { s } } , \boldsymbol { \epsilon } \right)=\mathbf { R } \left( - \mathbf { \theta } _ { \mathbf { s } } \right) \mathbf { E } ( \boldsymbol { \epsilon } ) \mathbf { R } \left( \boldsymbol { \theta } _ { s } \right)=\left( \begin{array} { c c } { \cos \left( \theta _ { s } \right) } & { \sin \left( \theta _ { s } \right) } \\ { - \sin \left( \theta _ { s } \right) } & { \cos \left( \theta _ { s } \right) } \end{array} \right) \left( \begin{array} { c c } { \frac { 1 } { 1 + \epsilon } } & { 0 } \\ { 0 } & { \frac { 1 } { 1 - \delta \epsilon } } \end{array} \right) \left( \begin{array} { c c } { \cos \left( \theta _ { s } \right) } & { - \sin \left( \theta _ { s } \right) } \\ { \sin \left( \theta _ { s } \right) } & { \cos \left( \theta _ { s } \right) } \end{array} \right)
	\end{equation}
	
	This is represented in figure S1b.
	
	We now consider the Moir\'e wavelengths for two lattices -- one of which is oriented along the k$_x$ direction and strained by percentage $\epsilon$ at an angle $\theta_s$ to the k$_x$ axis, and the other is rotated at angle $\theta$ relative to the first but is unstrained. 
	
	\begin{equation}
	\boldsymbol { k } _ { \mathbf { s1 } } = \boldsymbol {S} ( \boldsymbol { \theta_S,\epsilon } ) \boldsymbol { k } _ { \mathbf { 1 } } , \boldsymbol { k } _ { s2 } = \boldsymbol { S } ( \boldsymbol { \theta_s,\epsilon } ) \boldsymbol { k } _ { 2 } , \boldsymbol { k } _ { s3 }  = \boldsymbol { S } ( \boldsymbol { \theta_s,\epsilon } ) \boldsymbol { k } _ { 3 }
	\end{equation}
	
	\begin{equation}
	\boldsymbol { k } _ { \mathbf { 1 } } ^ { \prime } = \boldsymbol { R } ( \boldsymbol { \theta } ) \boldsymbol { k } _ { \mathbf { 1 } } , \boldsymbol { k } _ { 2 } ^ { \prime } = \boldsymbol { R } ( \boldsymbol { \theta } ) \boldsymbol { k } _ { 2 } , \boldsymbol { k } _ { 3 } ^ { \prime } = \boldsymbol { R } ( \boldsymbol { \theta } ) \boldsymbol { k } _ { 3 }
	\end{equation}
	
	In this case, the Moir\'e reciprocal wavelengths are different but can be found as in the strainless case, just treating each set individually as represented in figure S1b
	
	\begin{equation}
	\boldsymbol { K } _ { 1 s } = \boldsymbol { k } _ { 1 } ^ { \prime } - \boldsymbol { k } _ { 1 s } , \quad \boldsymbol { K } _ { 2 s } = \boldsymbol { k } _ { 2 } ^ { \prime } - \boldsymbol { k } _ { 2 s } , \quad \boldsymbol { K } _ { 3 s } = \boldsymbol { k } _ { 3 } ^ { \prime } - \boldsymbol { k } _ { 3 s }
	\end{equation}
	
	Finally, Fourier transforming back recovers the real space Moir\'e wavelengths in each of the 3 directions
	
	\begin{equation}
	\boldsymbol{\left | M _ { 1 } \right|} = \frac { ( 4 \pi ) } { \sqrt { 3 } \boldsymbol{\left| K _ { 1 s } \right| }} , \boldsymbol{\left| M _ { 2 } \right|} = \frac { ( 4 \pi ) } { \sqrt { 3 } \boldsymbol{\left| K _ { 2 s } \right| }} , \boldsymbol{\left| M _ { 3 } \right|} = \frac { ( 4 \pi ) } { \sqrt { 3 } \boldsymbol{\left| K _ { 3 s } \right| }}
	\end{equation}
	
	In the experiment, we start with the Moir\'e wavelengths which are measured in real space. We numerically solve for the three unknown parameters $\theta_T$, $\theta_s$ and $\epsilon$ that best fit to the three measured Moir\'e wavelengths.
	
	As an example, of the two samples near magic angle presented in this work, one had wavelengths of 13.72 nm, 12.7 nm and 10.18 nm. Running a numerical fit which generates $| \boldsymbol { M _ { 1 } } |$,$| \boldsymbol { M _ { 2 } } |$and$| \boldsymbol { M _ { 3 } } |$ and with optimization parameters of $\theta_T$, $\theta_s$ and $\epsilon$ using the derived relations, we can find the precise combination of parameters which produces the experimental conditions. For this case, the numerical fit gives $\theta_T$=1.152$^\circ$, $\theta_s$=25.5$^\circ$, and $\epsilon$=0.68\% with $| \boldsymbol { M _ { 1p } } |$=13.72,$| \boldsymbol { M _ { 2p } } |$=12.70 and $| \boldsymbol { M _ { 3p } } |$=10.18. 
	
	\begin{figure*}[t]
		\includegraphics[width=\linewidth]
		{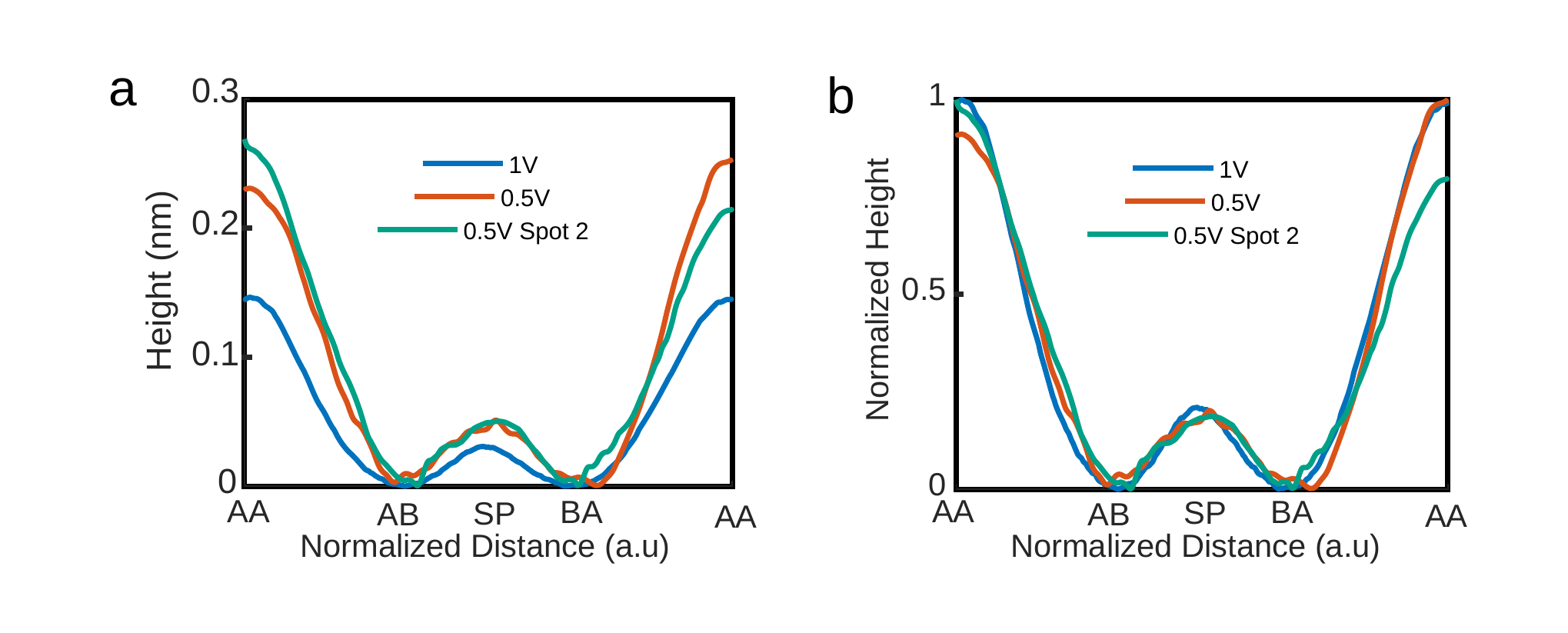}
		\caption{
			(a) Height profiles of 1.10$^\circ$ TBLG Moir\'e's at two STM biases and one different location. (b) Normalized height profiles of figure S2a.} 
	\end{figure*} 
	
	For the second magic angle sample, the wavelengths were 14.50, 13.20, and 10.84 nm and the numerical fit gives $\theta_T$=1.095$^\circ$, $\theta_s$=27$^\circ$, $\epsilon$=0.61\% with $| \boldsymbol { M _ { 1p } } |$=14.50 nm, $| \boldsymbol { M _ { 2p } } |$=13.20 nm and $| \boldsymbol { M _ { 3p } } |$=10.84 nm.
	
	We can also derive the area of the unit cells formed by these Moir\'e wavelengths to compare to angles derived in transport which do not know of local strain. The first example above would geometrically yield a triangle of area 61.7 nm$^2$. Without knowledge of strain, one would then assume this is an equilateral triangle and deduce a mean Moir\'e wavelength of 11.93 nm which would lead to an assumed twist angle of 1.18$^\circ$. For the second case, a similar treatment leads to a wavelength of 12.34 nm and an angle of 1.12$^\circ$, both very similar to the numbers obtained by our numerical model, however less accurate due to the lack of strain corrections.
	
	\section*{S2: Consistency of height profiles at different biases}
	STM topographic height receives contributions both from actual height variations in the sample as well as the integrated local density of states variations. We find that at bias setpoints at 0.5 V or larger (above the main features of the LDOS), the normalized height profiles across the sample are independent of bias. Figure S2a shows the unnormalized height profiles at two different setpoints and one different area of our 1.10$^\circ$ sample. There is a clear difference in absolute heights, but when normalized to the maximum height contrast between the AA and AB regions, the data collapse onto a 
	single curve. 
	
	\begin{figure*}[t]
		\includegraphics[width=\linewidth]
		{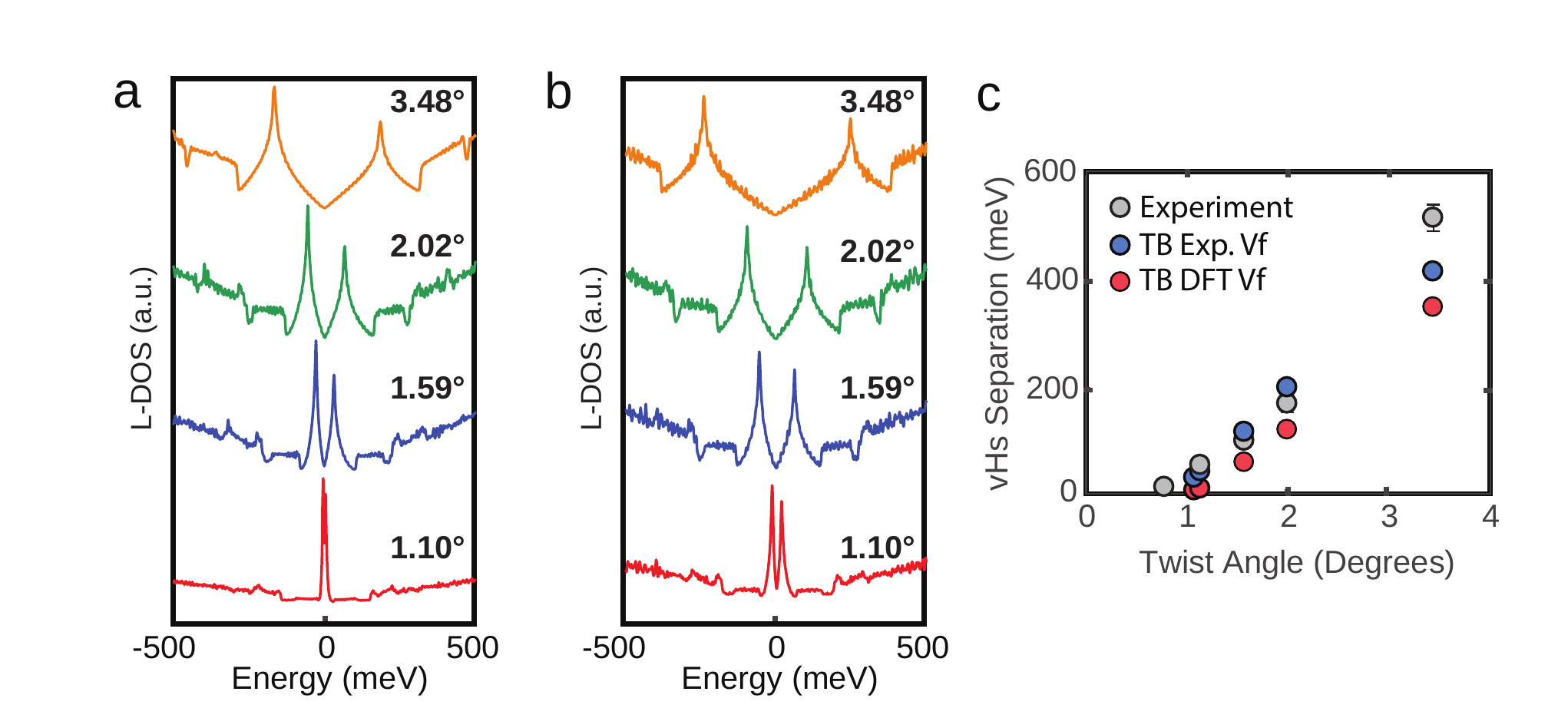}
		\caption{
			(a-b) Tight-binding calculations for TBLG LDOS using DFT-LDA (a) and experimental (b) monolayer graphene V$_f$ for various twist angles in this study. (c) A comparison of the vHs separations obtained with the two methods vs STS measurement.} 
	\end{figure*} 
	
	\section*{S3: LDOS calculations with different intralayer hopping parameters}
	
	In the main text, we show the LDOS calculation results obtained from a tight binding model with an enlarged intralayer hopping to include the many-body effects that determine the Fermi velocity of monolayer graphene correctly. Here, we also show results with all tight binding parameters fitted to DFT band structures, as shown in Fig. S3(a) and (b). The results with all DFT-fitted parameters (Fig.S3(a)) are consistent with previous published results with DFT or tight binding calculations with DFT fitted parameters, but generally underestimate the separation between vHs peaks compared with experimental results. In particular, the vHs peak separations for the system with twist angle at 1.10 degree is extremely small, with a value around 6meV. In contrast, the agreement with experimental results is significantly improved when we use an enlarged intralayer hopping that corresponds to the experimental Fermi velocity in monolayer graphene (see Fig. S3(c)).
	
	\section*{S4: Consideration of Experimental Artifacts in vHs separation}
	At the angle of 1.10$^\circ$, we measure an experimental vHs splitting of 55meV at zero doping. It is important to consider experimental artifacts at the single particle level that contribute to the measured splitting. One possibility is tip induced band bending. Indeed, previous STM measurements showed that the splitting between the vHs was dependent on doping in a manner consistent with tip induced band bending \cite{wong15}. However, these effects were primarily seen at large twist angles where the density of states is small. A direct extrapolation of the previous experiments shows that this effect should be negligible at the small angles measured here. A second possibility is the presence of displacement field in our experiment due to the presence of asymmetric gating conditions -- the STM tip is held several Angstroms away from the sample which is at ground potential, while the gate electrode is the silicon wafer that is approximately 1 micrometer away from the surface. For the spectra shown in figure 2a, the back gate is held at ground potential while the bias of the sample is swept through the vHs. The displacement field under this condition at the bias of the vHs is 2.5 x $10^{-5} V/\AA$. Tight-binding calculations performed for asymmetrically-doped layers indicate that this small value of the displacement field is negligible \cite{tram16}. A third possibility is the effect of the measured heterostrain in our samples. All of our samples and those of previous works show the presence of a small degree of heterostrain between the two graphene layers, with strain values varying between 0.1\% and 0.7\% for our measurements. For the 1.10$^\circ$ spectrum shown in figure 2a, the heterostrain is about 0.7\%. The effect of heterostrain on TBLG has been investigated using an ad-hoc tight binding model recently \cite{huder18} where it was predicted that uniform heterostrain between the layers leads to a third peak in the L-DOS at the Dirac point between the two vHs's. We have never observed such a third peak in any of our samples, and prior STM works of TBLG on hBN substrates have also not seen any evidence for this \cite{wong15,kim17,huang18}. While a full theory for heterostrained TBLG that accurately predicts LDOS spectra is yet to be established, we have several reasons to believe that the effect of strain on our spectroscopy and the vHs splitting is small: (a) our spectroscopy at all twist angles with variable degrees of small heterostrain show consistent features and trends between each other and previous works \cite{wong15,kim17,huang18} when strain was not considered but was also likely present to some degree. (b) Our tight-binding calculations with an experimental Fermi velocity accurately reproduce our experimental splittings. (c) Measurements on a second angle near the magic angle TBLG at a different heterostrain of 0.6\% shows similar splitting between the vHs.

	\section*{S5: Hartree-Fock Mean-Field Model}
	
	\begin{figure*}[t]
		\includegraphics[width=0.5\linewidth]
		{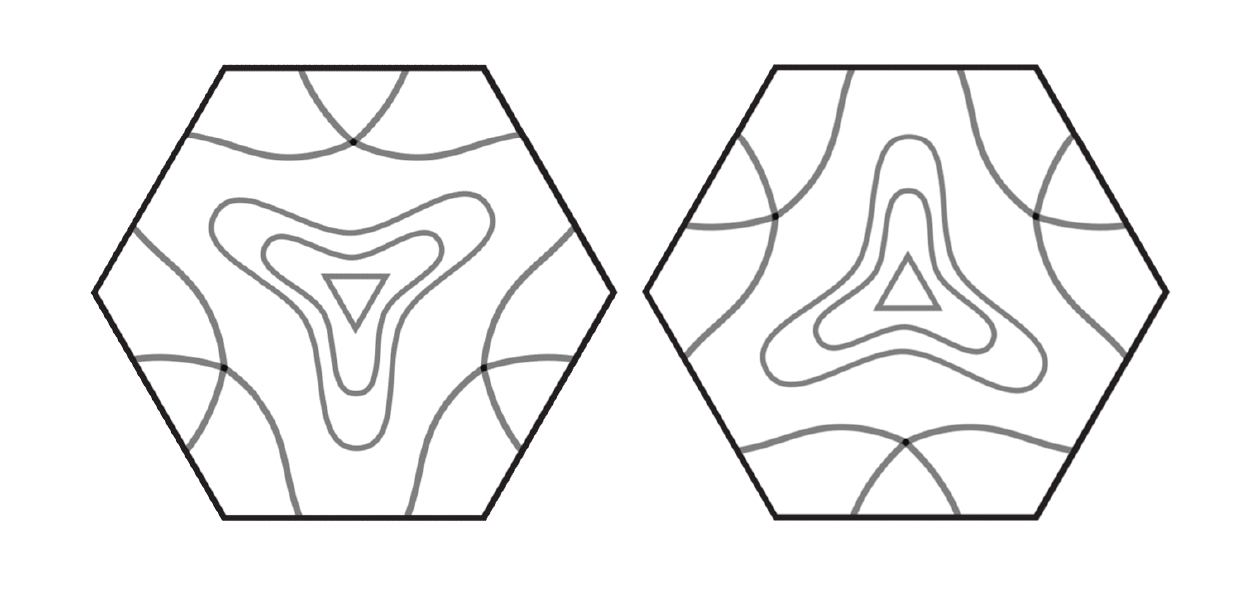}
		\caption{
			Contours of constant energy at small energies in the vicinity of the vHs at K and K’, following Ref 34. The Fermi surfaces show strong nesting. } 
	\end{figure*}

We perform a Hatree-Fock treatment of the Hamiltonian
\begin{equation}
H=\sum\limits_{\left\langle i,j\right\rangle}\sum\limits_{\sigma=\uparrow,\downarrow} \left[tc^\dagger_{i,\sigma}c_{j,\sigma}+{\rm H.c.}+V_1n_in_j\right] +\sum_i U n_{i,\uparrow}n_{i,\downarrow}
\end{equation}
where $\left\langle i,j\right\rangle$ are nearest neighbors on the Honeycomb lattice.
Around van Hove filling we decouple the Hartree term $U$ using the nesting vector in a self-consistent Hartree-Fock treatment. $V_1$ simply renormalizes the hopping 
$t\to t+\Sigma(\mu)$ (and with it the bandwidth) according to first order perturbation theory
\begin{equation}
\Sigma(\mu)=V_1\lim\limits_{\eta\to 0}\sum_{s=\pm}\int d\omega f(\omega,\mu)\int_{\vec k}  d\vec k \frac{s\left\langle i\right|\vec k\rangle\left\langle k\right|\vec j\rangle}{\omega -\epsilon_{s}(k,t)+i\eta},
\end{equation} 
where the integral over $\vec k$ runs over the Brillouin zone, $f(\omega,\mu)$ is the Fermi-Dirac distribution, $\left|\vec j\right\rangle$ and $\left|\vec k\right\rangle$ are the single particle wavefunctions in real and momentum space and $\epsilon_{s}(\vec k,t)$ is the dispersion relation of the bipartite lattice (the non-interacting part of the Hamiltonian ($U=V_1=0$) can be written $H^0=\sum_{s=\pm,\sigma\uparrow,\downarrow}\int_{\vec k}\epsilon_{s}(\vec k,t)c^\dagger_{k,\sigma}c_{k,\sigma} $). The density of states reads
\begin{equation}
{\text{DOS}}=-\frac{1}{\pi}\lim\limits_{\eta\to0}\sum_{s=\pm}\int_{\vec k}  d\vec k \; {\rm Im}\left[\frac{1}{\omega -\epsilon_{s}(\vec k,t)+i\eta}\right].
\end{equation}
At van-Hove filling $\mu=t$ the Fermi surface is perfectly nested by three nesting vectors $\vec Q_i$ (being rotated by $120^\circ$). We break the $C_3$ symmetry of the underlying lattice by picking one the three $\vec Q=\vec Q_1$.
Due to the strong enhancement around the van Hove filling this will open a gap by the self-consistent Hatree-Fock equation of 
\begin{equation}
\frac{1}{U}\stackrel{!}{=}\int_{\vec k}d\vec k\frac{1}{2\sqrt{\Delta^2+((\epsilon_+(\vec k,t)-\epsilon_+(\vec k-Q,t))/2)^2}}\left[f(E^-(\vec k),\mu)-f(E^+(\vec k),\mu)\right]
\end{equation} 
with $E^{\pm}(\vec k)=(\epsilon_+(\vec k,t)-\epsilon_+(\vec k-\vec Q,t))/2\pm\sqrt{\Delta^2+((\epsilon_+(\vec k,t)-\epsilon_+(\vec k-\vec Q,t))/2)^2}$.

	\section*{S6: C$_3$ Symmetry Breaking}
	In the experimental data we see a clear indication for symmetry breaking of the C$_3$ symmetry of the underlying lattice for states close to Fermi-surface. While strain can provide a direction that determines the electronic symmetry breaking, the underlying structure of the Fermi surface can determine the magnitude of electronic symmetry breaking. One possible way this can happen is via Fermi surface nesting. Considering the results given in Ref.~\cite{kim16},  we find a qualitative picture for the Fermi-surface as given in figure S4. The figure shows that the nesting vector connecting the nearly parallel surfaces between the one orbital degree of freedom and the other (left and right figure S4 respectively), stays approximately constant and points along the AA to nearest AA direction in real space like observed in the experiment. 
	
	\begin{figure*}[t]
		\includegraphics[width=\linewidth]
		{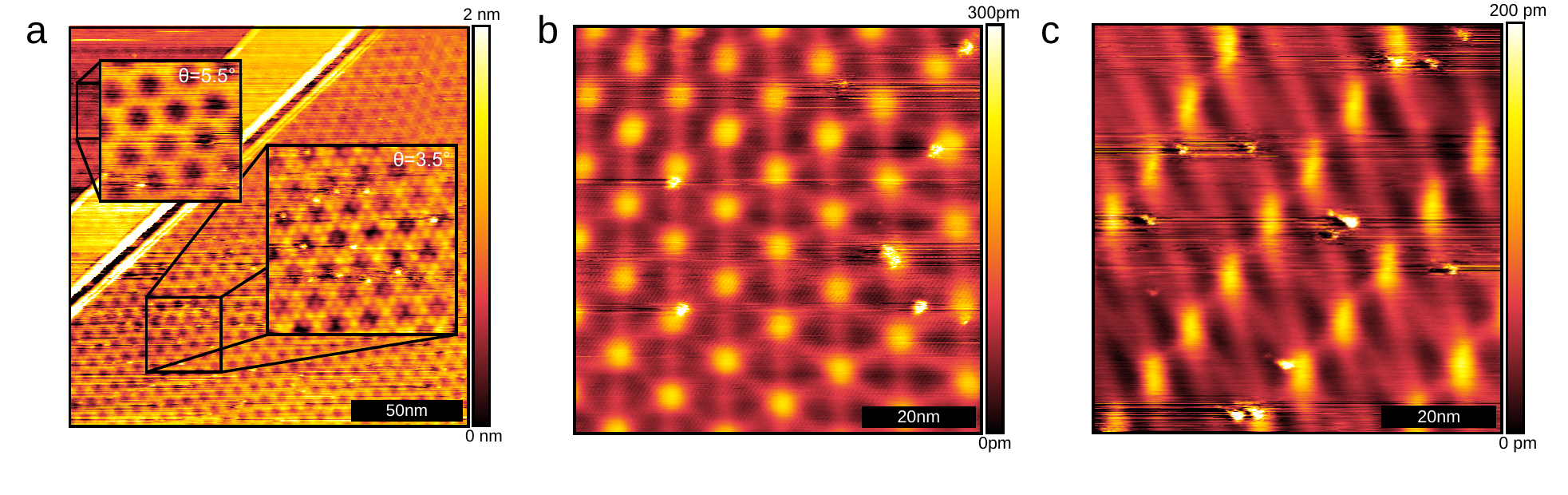}
		\caption{
			(a) STM topography where a wrinkle causes a change in the TBLG twist angle. (b) STM topography of an area where the twist angle spatially evolves with no structural fault. (c) STM topography of an area of extreme strain where the Moir\'e lattice is completely distorted.} 
	\end{figure*} 
	
	\section*{S7: Inhomogeneity Extremes of Samples}
	In the eight samples that we measured for this study, we observed inhomogeneity over large areas in all samples fabricated by this method (which also produced the superconducting samples in transport). Some extreme examples are shown in figure S5(a-c) such as wrinkles, spatially evolving twists and extreme strains.
	
	\bibliographystyle{apsrev4-1}
	\bibliography{tblgbibtexsupp}